\documentclass[runningheads]{llncs}
\usepackage[T1]{fontenc}    
\usepackage{hyperref}       
\usepackage{url}            
\usepackage{booktabs}       
\usepackage{amsfonts}       
\usepackage{nicefrac}       
\usepackage{microtype}      

\usepackage{amsmath,amssymb}
\usepackage{comment}
\usepackage{color}
\usepackage{graphicx}
\usepackage{algorithm,algcompatible,algpseudocode}
\usepackage{multirow}
\usepackage{pbox}

\DeclareMathOperator*{\argmin}{args\,min}

\begin{document}

\title{
Progressive Subsampling for Oversampled Data - Application to Quantitative MRI
}
\titlerunning{
Progressive Subsampling for Oversampled Data - Application to qMRI
}

\author{
Stefano B. Blumberg\inst{1}\thanks{Corresponding Author} \and Hongxiang Lin\inst{1,3} \and Francesco Grussu\inst{1,2} \and \\
Yukun Zhou\inst{1} \and Matteo Figini\inst{1} \and Daniel C. Alexander\inst{1} 
}

%
\authorrunning{
Stefano B. Blumberg et al.
}
%
\institute{
    University College London (UCL) \and 
    Vall d’Hebron Barcelona Hospital \and 
    Zhejiang Lab \\
    \email{
        hxlin@zhejianglab.edu.cn, stefano.blumberg.17@ucl.ac.uk
    }
}

\maketitle

\begin{abstract}

We present PROSUB: PROgressive SUBsampling, a deep learning based, automated methodology that subsamples an oversampled data set (e.g. channels of multi-channeled 3D images) with minimal loss of information.  We build upon a state-of-the-art dual-network approach that won the MICCAI MUlti-DIffusion (MUDI) quantitative MRI (qMRI) measurement sampling-reconstruction challenge, but suffers from deep learning training instability, by subsampling with a hard decision boundary.  PROSUB uses the paradigm of recursive feature elimination (RFE) and progressively subsamples measurements during deep learning training, improving optimization stability.  PROSUB also integrates a neural architecture search (NAS) paradigm, allowing the network architecture hyperparameters to respond to the subsampling process.  We show PROSUB outperforms the winner of the MUDI MICCAI challenge, producing large improvements >$18 \% $ MSE on the MUDI challenge sub-tasks and qualitative improvements on downstream processes useful for clinical applications.  We also show the benefits of incorporating NAS and analyze the effect of PROSUB's components.  As our method generalizes beyond MRI measurement selection-reconstruction, to problems that subsample and reconstruct multi-channeled data, our code is \cite{ourcodegithub}.

\keywords{Magnetic Resonance Imaging (MRI) Protocol Design, Recursive Feature Elimination, Neural Architecture Search}
\end{abstract}

\section{Introduction}

Multi-modal medical imaging gives unprecedented insight into the microstructural composition of living tissues, and provides non-invasive biomarkers that hold promise in several clinical contexts.  In particular, quantitative MRI fits a model in each pixel of a multi-channel acquisition consisting of multiple images each with unique contrast obtained by varying multiple MRI acquisition parameters, see e.g. \cite{hutter2018}.  This provides pixel-wise estimates of biophysical tissue properties \cite{grussu2020multi}. In spite of this potential, comprehensively sampling high-dimensional acquisition spaces leads to prohibitively long acquisition times, which is a key barrier to more widespread adoption of qMRI in clinical use.
\\
\indent The MUlti-DIffusion (MUDI) MRI challenge \cite{mudichallengeweb,pizzolato2019} addressed this by providing data covering a densely-sampled MRI acquisition space (3D brain images with $ 1344 $ channels).  The task was to reconstruct the full set of measurements from participant-chosen measurements from a small subsample, i.e. to obtain economical, but maximally informative acquisition protocols for any model that the full data set supports. That involves two sub-tasks: selecting the most informative measurements, and reconstructing the full data set from them.  The challenge winner was SARDU-Net \cite{grussu2020,grussu2021,pizzolato2019} with a dual-network strategy, that respectively subsamples the measurements, then reconstructs the full dataset from the subsampled data.  However, SARDU-Net selects different sets of measurements with a hard decision boundary on each training batch, altering the second network's input across different batches.  This can cause instability, suupl. mat. fig.~\ref{sardunet_instability:fig} and \cite{blumberg2019,karras2018} show similar issues produce training instability.  Furthermore, the popularity of paradigms such as recursive feature elimination (RFE), suggests that subsampling all of the measurements required immediately, is suboptimal.  These two issues may lead to substandard performance.
\\
\indent We propose PROSUB, a novel automated methodology that selects then reconstructs measurements from oversampled data.  Unlike classical approaches to experiment design \cite{alexander2008}, we approach the MUDI challenge in a new model-independent way.  PROSUB builds upon the SARDU-Net by (i) using a form of RFE which progressively removes measurements across successive RFE steps and (ii) learning an average measurement score across RFE steps, which chooses the measurements to remove or preserve. This enhances the stability of our optimization procedure. Within each RFE step, PROSUB (iii) progressively subsamples the required measurements during deep learning training, building upon \cite{blumberg2019,karras2018}, that improves training stability. Also, PROSUB (iv) incorporates a generic neural architecture search (NAS) paradigm in concurrence to the RFE – so the architectures may respond to the measurement subsampling process.
\\
\indent Our implementation \cite{ourcodegithub} is based on AutoKeras \cite{jin2019}, KerasTuner \cite{omalley2019}.  PROSUB outperforms the SARDU-Net and SARDU-Net with AutoKeras NAS by $ >18 \% $ MSE on the publicly available MUDI challenge data \cite{mudidata,pizzolato2019}.  We show qualitative improvements on downstream processes: T2\textsuperscript{*},FA,T1,Tractography useful in clinical applications \cite{andica2020,deoni2010,granziera2021,henderson2020,lehericy2020}.  We examine the effect of how PROSUB's components, including NAS, improve performance.  We release the code \cite{ourcodegithub} as PROSUB is not limited to subsampling MRI data for microstructure imaging.

\section{Related Work and Preliminaries}\label{preliminaries_and_related_work:sec}

\textbf{Problem Setting} Suppose we have an oversampled dataset $ \textbf{x} = \{ \textbf{x}_{1}, ..., \textbf{x}_{n} \} \in \mathbb{R}^{n \times N} $ where each sample has $ N $ measurements $ \textbf{x}_{i} \in \mathbb{R}^{N} $.  The aim is to subsample $ M < N $ measurements  $ \widetilde{\textbf{x}} = \{ {\widetilde{\textbf{x}}_1, ..., \widetilde{\textbf{x}}_n} \} \in \mathbb{R}^{n \times M}, \ \widetilde{\textbf{x}}_{i} \in \mathbb{R}^{M} $, with the same M elements of each $ \textbf{x}_{i} $  in each $ \widetilde{\textbf{x}}_{i} $.  We aim to lose as little information as possible when choosing $ \widetilde{\textbf{x}} $, thereby enabling the best recovery of the full data set $ \textbf{x} $.  We therefore have two interconnected problems i) choosing which measurements to subsample,  ii) reconstructing the original measurements from the subsampled measurements.  We achieve this by (i) constructing a binary mask $ m $ containing $ M $ ones and $ N-M $ zeros so $ \widetilde{\textbf{x}} = m \cdot \textbf{x} $, (ii) with a neural network $ \mathcal{R} $.
\\
\\ \textbf{SARDU-Net and Dual-Network Approaches} The SARDU-Net \cite{grussu2020,grussu2021,pizzolato2019}, which is used for model-free quantitative MRI protocol design and won the MUDI challenge \cite{pizzolato2019}, has two stacked neural networks, trained in unison.  The first network learns weight $ w $ from $ \textbf{x} $.  $ N - M $ smallest values of $ w $ are clamped to $ 0 $ and the first network subsamples and selects the measurements, by outputting $ \textbf{x} \cdot w $.  The second network then predicts the original data from $ \textbf{x} \cdot w $.  Related dual-network approaches include \cite{dovrat2019}, which processed large point clouds, differing from our problem, as we do not assume our data has a spatial structure.  We build upon the SARDU-Net and we use it as a baseline in our experiments. 
\\
\\ \textbf{Recursive Feature Elimination (RFE)} One of the most common paradigms for feature selection is RFE, which has a long history in machine learning \cite{scikitlearnrfe,zheng2018}.  Recursively over steps $ t = 1,... , T $, RFE prunes the least important features based on some task-specific importance score, successively analyzing less and less features over successive steps.  We use a form of RFE in PROSUB.
\\
\\ \textbf{Neural Architecture Search (NAS)} Selecting neural network architecture hyperparameters e.g. number of layers and hidden units, is a task-data-dependent problem, where the most common strategies are random search~\cite{bergstra2012} or grid search~\cite{larochelle2007}. NAS approaches, see e.g. \cite{elsken2019}, outperform classical approaches with respect to time required to obtain high-performing models and can broadly be seen as a subfield of Automated Machine Learning (AutoML) \cite{hutter2019}.  PROSUB uses a generic NAS paradigm which optimizes network architectures over successive steps $ t = 1,... , T $ in an outer loop.  In an inner loop, with fixed architecture (and fixed $ t $), we perform standard deep learning training across epochs $ e = 1,..., E $, caching network training and validation performance $ r_{t}^{e} $ after each epoch.  At the end of step $ t $, the previous losses $ \{ r^{i}_{j} : i \leq E, j \leq t \} $ are used to update the network architectures for step $ t + 1 $.  Our implementation is based on AutoKeras \cite{jin2019}, with KerasTuner \cite{omalley2019}, which has good documentation and functionality.

\section{Methods}

\begin{figure}[t]
\centering
\includegraphics[width=1\linewidth]{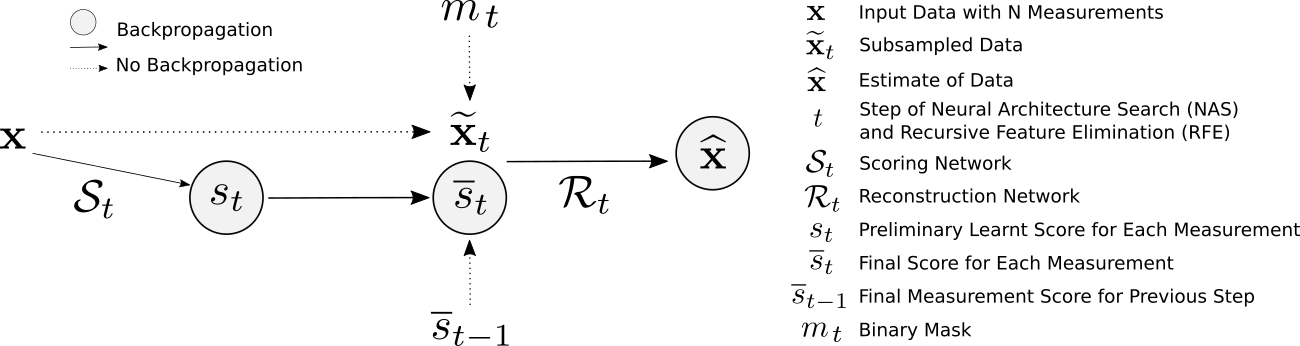}
\caption{
The computational graph of PROSUB.}\label{computational_graph:fig}
\end{figure}

We address the interdependency of the sampling-reconstruction problem with a dual-network strategy in section~\ref{scoring_reconstruction_networks:sec}, illustrated in fig.~\ref{computational_graph:fig}.  In section~\ref{subsampling_measurements:sec} we progressively construct our mask $ m $, used to subsample the measurements.  PROSUB has an outer loop: steps $ t = 1,..., T $ where we simultaneously perform NAS and RFE, choosing the measurements to remove via a score, averaged across the steps, whilst simultaneously updating the network architecture hyperparameters.  For fixed $ t $, we perform deep learning training as an inner loop across epochs $ e = 1,..., E $, where we learn the aforementioned score and also progressively subsample the measurements.  We summarize PROSUB in algorithm~\ref{main:alg}.

\subsection{Scoring-Reconstruction Networks}\label{scoring_reconstruction_networks:sec}

Inspired by \cite{dovrat2019,grussu2020}, we use two neural networks $ \mathcal{S}_{t},\mathcal{R}_{t} $, trained in unison, to address the interdependency of our two interconnected problems.
\\
\\ \textbf{Scoring Network} The first network $ \mathcal{S}_{t} $ learns a preliminary score with a sigmoid activation in its last layer, to weight each measurement:
\begin{equation}\label{scoring_network:eq}
    s_{t} = \mathcal{S}_{t}(\textbf{x}) \ \ \ s_{t} \in (0,2)^{n \times N}.
\end{equation}

\noindent \textbf{Mask} As described in section~\ref{preliminaries_and_related_work:sec}, we use an array $ m_{t} \in [0,1]^{N} $ as a mask
\begin{equation}\label{mask:eq}
    \widetilde{\textbf{x}}_{t} = m_{t} \cdot \textbf{x}
\end{equation}
to subsample the measurements.  We describe in section~\ref{subsampling_measurements:sec}, how we progressively and manually set $ m_{t} $ to have $ N-M $ entries with $ 0 $.
\\
\\ \textbf{Average Measurement Score} To score each measurement in step $ t $, we use an exponential moving average, commonly used in time-series analysis (e.g \cite{hamilton1994}),  across the scores of previous steps $ \overline{s}_{1},..., \overline{s}_{t-1} $ and $ s_{t} $, to obtain a better estimate of the score and reduce the effect of the current learnt score $ \sigma_{t}$, if network performance is poor.  With moving average coefficient hyperparameter $ \alpha_{t} $ we calculate
\begin{equation}\label{average_measurement_score:eq}
   \overline{s}_{t} = \alpha_{t} \cdot s_{t} + (1 - \alpha_{t}) \cdot \overline{s}_{t-1}
\end{equation}
and we use $ \alpha_{t} = \frac{T-t}{T-1} $ . The averaged score $ \overline{s}_{t} $ is used to weight the subsampled measurements and to construct the mask (described in section~\ref{subsampling_measurements:sec}).  
\\
\\ \textbf{Reconstruction Network}  The second network $ \mathcal{R}_{t} $ takes the weighted subsampled measurements to estimate $ \textbf{x} $ with $ \widehat{\textbf{x}} $, then passed through $ Loss $ (we use MSE)
\begin{equation}\label{reconstruction_forward_pass:eq}
    L = Loss(\widehat{\textbf{x}}, \textbf{x}), \ \ \ \widehat{\textbf{x}} =  \mathcal{R}_{t}(\widetilde{\textbf{x}}_{t} \cdot \overline{s}_{t})
\end{equation}
and the gradients from $ L $ are then backpropagated through $ \mathcal{R}_{t}, \mathcal{S}_{t} $.

\subsection{Constructing the Mask to Subsample the Measurements}\label{subsampling_measurements:sec}

\begin{algorithm}[t] 
\caption{PROSUB: PROgressive SUBsampling for Oversampled Data}\label{main:alg}
\label{alg:main}
\textbf{Data and Task:} $ \textbf{x} = \{\textbf{x}_{1}, ..., \textbf{x}_{n} \}, \ \textbf{x}_{i} \in \mathbb{R}^{N}, \ \ M < N $
\\ \textbf{Training and NAS:} $ 1 \leq E_{d} \leq E, \ 1 < T_{1} < T, \ NAS \gets AutoKeras $
\\ \textbf{Scoring and RFE:} $ \alpha_{t} \gets \frac{T-t}{T-1}, \ \ D_{t} \gets \approx \frac{N-M}{T-T_{1}+1} $
\\ \textbf{Initialize:} $ m_{1} \gets [1]^{N}, \ \overline{s}_{0} \gets [0]^{N} $
\begin{algorithmic}[1]
\For{$ t \gets 1, ..., T_{1}, ..., T $}  \Comment{RFE and NAS steps}\label{RFE_and_NAS_steps:line}
    \If{$1 \leq t < T_1 $}
        \State{$ D = \emptyset $ \Comment{No measurements to subsample}}\label{no_measurements_subsample:line}
    \ElsIf{$ T_{1} \leq t \leq T $} \Comment{Subsampling stage} \label{subsampling_stage:line}
        \State{$ D = \underset{{\scriptscriptstyle j=1, ..., D_{t} }}{{argsmin}} \{ \overline{s}_{t-1}[j] : m_{t}[j] = 1 \} $} \Comment{Measurements to subsample Eq.~\ref{choose_measurements_subsample:fig}}\label{choose_measurements_subsample:line}
    \EndIf
    \For{$ e \gets 1, ..., E_{d}, ..., E $ } \Comment{Training and validation epoch}\label{deep_learning_training:line}
        \State{$ m^{e}_{t} \gets \max \{m_{t} - \frac{(e-E_{d})\mathbb{I}_{e \geq E_{d}}}{E_{d}} \cdot \mathbb{I}_{i \in D}(i), 0 \}  $} \Comment{Compute mask Eq.~\ref{removing_measurements_during_training:eq}}\label{removing_measurements_during_training:line}
        \State{$ s_{t} = \mathcal{S}(\textbf{x}), \ \ \widetilde{\textbf{x}}_{t} = m^{e}_{t} \cdot \textbf{x}_{t} $} \Comment{Forward pass Eq.~\ref{scoring_network:eq},\ref{mask:eq}}
        \State{$ \overline{s}_{t} = \alpha_{t} \cdot s_{t} + (1 - \alpha_{t}) \cdot \overline{s}_{t-1} $} \Comment{Average measurement score Eq.~\ref{average_measurement_score:eq}}\label{average_measurement_score:line}
        \State{$ r^{e}_{t} \gets L(\widehat{\textbf{x}},\textbf{x}), \ \ L =  L(\widehat{\textbf{x}},\textbf{x}), \ \ \widehat{\textbf{x}} = \mathcal{R}_{t}(\widetilde{\textbf{x}} \cdot \overline{s}_{t}) $  \Comment{Forward/backward pass Eq.\ref{reconstruction_forward_pass:eq} }}
        \EndFor
    \State{Use NAS, $ \{ r^{i}_{j} : i \leq E, j \leq t \} $, to calculate $ \mathcal{R}_{t+1},\mathcal{S}_{t+1} $ } \Comment{Update architectures}
    \State{$ m_{t+1} \gets m^{E}_{t}, \ cache \ \overline{s}_{t} $}
\EndFor
\State \Return{$ m_{T} $, $ \overline{s}_{T} $, $ \mathcal{R}_{T} $ -- use as described in section~\ref{preliminaries_and_related_work:sec}}
\end{algorithmic}
\end{algorithm}

We construct a mask $ m_{t}^{e} $, used to subsample the measurements in section~\ref{preliminaries_and_related_work:sec} and eq.~\ref{mask:eq}.  We progressively set $ N-M $ entries of $ m_{t}^{e} $ to zero across NAS and RFE outer loop $ t=1,..., T $ and deep learning inner loop $ e=1,..., E $. We refer to algorithm~\ref{alg:main} for clarity.
\\
\\ \textbf{Outer Loop: Choosing the Measurements to Remove}  Following standard practise in RFE e.g. \cite{scikitlearnrfe,zheng2018}, we remove the measurements recursively, in our case, across steps $ t=1,..., T $ in alg.-line~\ref{RFE_and_NAS_steps:line}.  We split the RFE in two stages, by choosing a dividing step, hyperparameter $ 1 < T_{1} < T $.
\\ \indent In the first stage $ t=1,..., T_{1} $ the optimization procedure learns scores $ \overline{s}_{t} $ and optimizes the network architectures via NAS.  In alg.-line~\ref{no_measurements_subsample:line},  we choose no measurements to subsample ($ D = \emptyset $) thus the mask $ m_{t}^{e} = m_{t} $ is fixed in alg.-line~\ref{removing_measurements_during_training:line}. 
\\ \indent In the second stage $ t = T_{1}, ..., T $, we perform standard RFE.  We first choose a hyperparameter $ D_{t} \in \mathbb{N} $ -- the number of measurements to subsample in step $ t $.  In this paper, we remove the same number of measurements per step, so $ D_{t} \approx \frac{N-M}{T-T_{1}+1} $.  In alg.-line~\ref{choose_measurements_subsample:line} we then choose the measurements to remove in RFE step $ t $, which correspond to those with the lowest scores in the previous step
\begin{equation}\label{choose_measurements_subsample:fig}
  D = \argmin_{j=1, ..., D_{t}}  \{ \overline{s}_{t-1}[j] : m_{t}[j] = 1 \}, \ \ \ m_{t} \in \{0,1\}^{N}
\end{equation}
where here $ m_{t} $ indicates whether the measurement has been removed in previous steps $ < t $.  Our rationale is since the subsampled measurements are weighted by the score, used as inputs to the reconstruction network in eq.~\ref{reconstruction_forward_pass:eq}, setting lowest-scored values to $ 0 $ may have small effect on the performance (in eq.~\ref{reconstruction_forward_pass:eq}).
\\
\\ \textbf{Inner Loop: Progressively Subsampling the Measurements by Altering the Mask During Training} Given $ D $ -- computed in the outer loop alg.-line~\ref{choose_measurements_subsample:line} we progressively, manually, alter the mask $ m_{t}^{e} $ in the inner loop of deep learning training alg.-line~\ref{deep_learning_training:line}, i.e. gradually setting the value of these measurements to $ 0 $ in $ \widetilde{\textbf{x}}_{t} $.  We are inspired by \cite{blumberg2019,karras2018} which used a similar approach to improve training stability.  We alter $ m_{t}^{e} $ across chosen epochs $ e=E_{d},..., 2 \cdot E_{d} - 1 \leq E $ for hyperparameter $ E_{d} < \frac{E}{2} $, used in alg.-line~\ref{removing_measurements_during_training:line}, for indicator function $ \mathbb{I} $:
\begin{equation}\label{removing_measurements_during_training:eq}
    m^{e}_{t} = \max{\{ m_{t} - \frac{(e-E_{d})\mathbb{I}_{e \geq E_{d}}}{E_{d}} \cdot \mathbb{I}_{i \in D}(i), 0  \} }.
\end{equation}

\section{Experiments and Results}

\begin{figure}[H]
\centering
\includegraphics[width=\linewidth]{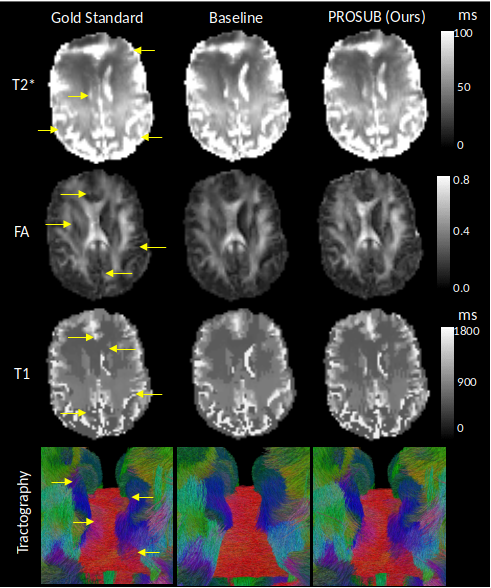} \\
\caption{
Qualitative comparison of downstream processes with useful clinical applications \cite{andica2020,deoni2010,granziera2021,henderson2020,lehericy2020}, of reconstructed $ N=1344 $ measurements from chosen $ M = 50 $ samples, on a random test subject.  PROSUB results are visually closer to the gold standard than the baseline. 
As MUDI data provides combined diffusion and relaxometry information, to evaluate the practical impact of the different reconstructions we estimated T1, T2* values and DTI parameters with a dictionary-based approach in \cite{garyfallidis2014}.
We show whole-brain probabilistic tractography examining reconstructed fibre tracts, colors correspond to direction, on multi-shell/tissue constrained spherical deconvolution via iFOD2 in \cite{jeurissen2014}.  Quantative improvements are, Baseline - Ours MSE, e.g. FA: 0.006 - 0.004, on NODDI parametric maps \cite{zhang2012}: 0.022 - 0.007 FICVF, 0.023 - 0.005 FISO, 0.020 - 0.013 ODI.
}
\label{brain_images:fig}
\end{figure}

\begin{figure}[ht]
\centering
\includegraphics[width=0.47\linewidth]{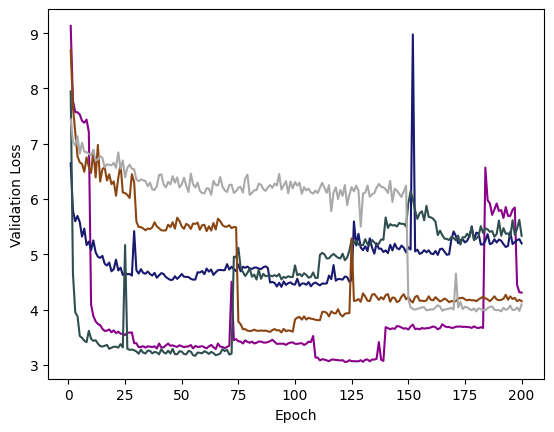}
\includegraphics[width=0.47\linewidth]{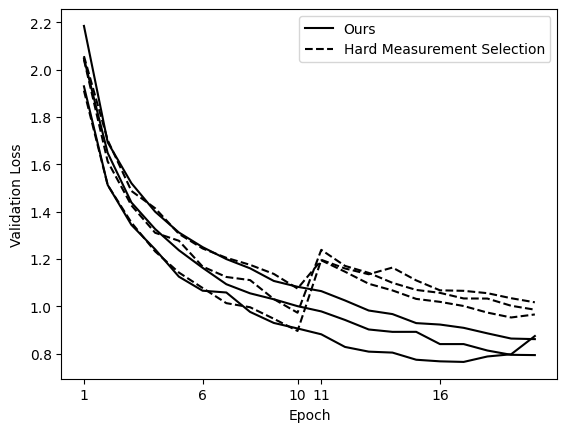}
\caption{ Left: Training losses of the SARDU-Net $ M=10 $ for different seeds/initializations (colors), the SARDU-Net selects different sets of measurements with a hard decision boundary on each training batch, altering the second network's input across different batches, producing instability. Right: Progressively subsampling measurements during training eq.~\ref{removing_measurements_during_training:eq}, compared to hard measurement selection (table~\ref{ablation:table}-row-3).
}\label{sardunet_instability:fig}\label{removing_measurements_during_training:fig}
\end{figure}

\noindent \textbf{MUDI Dataset and Task} Data of images from 5 subjects are from the MUDI challenge \cite{mudichallengeweb,pizzolato2019}, publicly available at \cite{mudidata}.  Data features a variety of diffusion and relaxometry (i.e. T1 and T2*) contrasts, and were acquired with the ZEBRA MRI technique \cite{hutter2018}.  The total acquisition time for these oversampled data sets was $ \approx 1h $, corresponding to the acquisition of $ N = 1344 $ measurements in this dense parameter space, resulting in $ 5 $, 3D brain images with $ 1344 $ channels (here unique diffusion- T2* and T1- weighting contrasts), with $ n \approx 558K $ brain voxels.  Detailed information is in \cite{hutter2018,pizzolato2019}.  We used the same task as the MICCAI MUDI challenge \cite{pizzolato2019}, where the participants were asked to find the most informative subsets of size $ M = 500,250,100,50 $ out of $ N $, while also estimating the fully-sampled signals from each of these subsets, and the evaluation is MRI signal prediction MSE.  The winner of the original challenge \cite{mudichallengeweb,pizzolato2019} was the aforementioned SARDU-Net \cite{grussu2020,pizzolato2019}.  In this paper, we also consider smaller subsets $ M = 40,30,20,10 $. 
\\ 
\\ \textbf{Experimental Settings} We did five-fold cross validation using two separate subjects for validation and testing.  We compare PROSUB and PROSUB w/o NAS with four baselines:  i) SARDU-Net-v1: winner of the MUDI challenge \cite{grussu2020,pizzolato2019}; ii) SARDU-Net-v2: latest official implementation of (i) \cite{grussuCode,grussu2021}; iii) SARDU-Net-v2-BOF: five runs of (ii) with different initializations, choosing the best model from the validation set;  iv) SARDU-Net-v2-NAS: integrating (ii) with AutoKeras NAS.  To reduce total computational time with NAS techniques, we performed all of the tasks in succession.  We first use algorithm~\ref{main:alg} with $ T_{1},T,M = 4,8,500 $, then take the final model, as initialization for algorithm~\ref{main:alg} with  $ T_{1},T,M = 1,5,250 $, performing this recursively for $ M=100,50,40,30,20,10 $, using the best model for each different $ M $.  Consequently, SARDU-Net-v2-BOF and the NAS techniques in table~\ref{mudi_crossval_test:table} are trained for approximately the same number of epochs.  We performed a brief search for NAS hyperparameters.  In figs.~\ref{ablation:table},\ref{hyperparameters_all:table} we examined the effect of PROSUB's components and present all hyperparameters.
\\ 
\\ \textbf{Main Results}  We present quantitative results in 
\begin{table}[t]
\centering
\caption{Whole brain Mean-Squared-Error between $ N=1344 $ reconstructed measurements and $ N $ ground-truth measurements, on leave-one-out cross validation on five MUlti-DIffusion (MUDI) challenge subjects.  The SARDU-Net won the MUDI challenge.
}\label{mudi_crossval_test:table}
\resizebox{\linewidth}{!}{
\begin{tabular}{c c c c c c c c c c}
\multicolumn{1}{c}{\multirow{2}{*}{\pbox{2cm}{}}} & &
\multicolumn{1}{c}{\multirow{2}{*}{\pbox{2cm}{}}} & 
\multicolumn{7}{c}{MUDI Challenge Subsamples M for $ N=1344 $}  \\
\cline{3-10}
                     & & & 500 & & 250 & & 100 & & 50 \\ \hline
SARDU-Net-v1 \cite{grussu2020,pizzolato2019} & Baseline &  &  $ 1.45 \pm 0.14 $  & &  $ 1.72 \pm 0.15 $ & & $ 4.73 \pm 0.57 $ & & $  5.15 \pm 0.63 $ \\
SARDU-Net-v2 \cite{grussuCode,grussu2021} & Baseline &  &  $  0.88 \pm 0.10 $  & &  $ 0.89 \pm 0.01 $ & & $ 1.36 \pm 0.14 $ & & $  1.66 \pm 0.10 $ \\
SARDU-Net-v2-BOF \cite{grussuCode,grussu2021} & Baseline &  &  $ 0.83 \pm 0.10 $  & &  $ 0.86 \pm 0.10 $ & & $ 1.30 \pm 0.12 $ & & $  1.67 \pm 0.12 $ \\
SARDU-Net-v2-NAS & Baseline  &  &  $ 0.82 \pm 0.13 $  & & $ 0.99 \pm 0.12 $ & &  $ 1.34 \pm 0.26 $ & & $ 1.76 \pm 0.24 $ \\ 
\hline
PROSUB w/o NAS & Ours & & $ 0.66 \pm 0.08 $ & & $ 0.67 \pm 0.09 $ & &  $ \textbf{0.88} \pm 0.07 $ & & $ 1.54 \pm 0.11 $ \\ 
PROSUB      & Ours  & & $ \textbf{0.49} \pm 0.07 $ & & $ \textbf{0.61} \pm 0.11 $ & & $ 0.89 \pm 0.11 $ & & $ \textbf{1.35} \pm 0.11 $ \\ 
\\
                    & &  & M = 40 & & 30 & & 20 & & 10 \\ \hline
SARDU-Net-v1 \cite{grussu2020,pizzolato2019} & Baseline &  &  $ 6.10 \pm 0.79 $  & &  $ 21.0 \pm 6.07 $ & & $ 19.8 \pm 9.26 $ & & $ 22.8 \pm 6.57  $ \\
SARDU-Net-v2 \cite{grussuCode,grussu2021} & Baseline & &  $ 1.95 \pm 0.12 $  & & $ 2.27 \pm 0.20 $ & & $ 3.01 \pm 0.45 $ & & $ 4.41 \pm 1.39 $ \\
SARDU-Net-v2-BOF \cite{grussuCode,grussu2021} & Baseline & &  $ 1.86 \pm 0.18 $  & & $ 2.15 \pm 0.23 $ & & $ 2.61 \pm 0.24 $ & & $ 3.74 \pm 0.66 $ \\
SARDU-Net-v2-NAS & Baseline  &  & $ 2.23 \pm 0.22 $ & & $ 6.00 \pm 7.14 $ & &  $ 2.82 \pm 0.41 $ & & $ 4.27 \pm 1.66 $ \\ 
\hline
PROSUB w/o NAS & Ours & & $ 1.81 \pm 0.18 $ & & $ 2.18 \pm 0.17 $ & & $ 2.72 \pm 0.34 $ & &  $ 3.91 \pm 0.22 $ \\ 
PROSUB      & Ours & &  $ \textbf{1.53} \pm 0.05 $ & & $ \textbf{1.87} \pm 0.19 $ & & $ \textbf{2.50} \pm 0.40 $ & & $ \textbf{3.48} \pm 0.55 $ \\ 
\end{tabular}
}
\end{table}
table~\ref{mudi_crossval_test:table} and note PROSUB's large improvements  >$18 \% $ MSE over all four baselines on the MUDI challenge sub-tasks. Using the Wilcoxon one-sided signed-rank test, a non-parametric statistical test comparing paired brain samples, our methods improvements have p-values of  $7.14$E$-08$, $9.29$E$-07$, $3.20$E$-06$, $9.29$E$-07$ over the four respective baselines with Bonferroni correction, thus are statistically significant.  We provide qualitative comparisons on a random test subject in fig.~\ref{brain_images:fig} on downstream processes (T2\textsuperscript{*},FA,T1,Tractography), useful in clinical applications \cite{andica2020,deoni2010,granziera2021,henderson2020,lehericy2020}.
\\
\\ \textbf{Discussion}  Without explicitly optimizing PROSUB's network architecture and training hyperparameters (fixing PROSUB's hyperparameters to the SARDU-Net hyperparameters at $ M = 500 $), the PROSUB still outperforms all SARDU-Net baselines for MUDI Challenge $ M $.  Concerning NAS, we note the SARDU-Net-v2-NAS generally underperforms the SARDU-Net w/o NAS.  Examining network performance during NAS (e.g. fig.~\ref{sardunet_instability:fig}), this is due to SARDU-Net performance being unstable due to its hard measurement selection, thus its performance is vulnerable to changes in architecture.  Passing poor results to the NAS then reduces the effectiveness of the NAS in identifying high-performing architectures for small $ M $. In contrast, PROSUB's progressive subsampling allows the NAS to identify better architectures than the SARDU-Net-v2-NAS.  PROSUB also outperforms the PROSUB w/o NAS i.e. the NAS is able to identify better performing architectures than the architecture chosen by the SARDU-Net for $ M = 500 $.  In tables~\ref{ablation:table},\ref{max_size:table},\ref{network_sizes:table},  we analyze the effect of PROSUB's components, note removing any of the three non-NAS contributions worsens performance;  we also show increasing network capacity in the PROSUB does not necessarily improve performance; we also list the architectures chosen by the NAS for SARDU-Net-v2-NAS and PROSUB.

\section{Future Work}

In future work, we could add an additional cost function to address the cost of obtaining specific combination of measurements from sets of MRI acquisition parameters; and develop a novel NAS algorithm to account for the concurrence of the subsampling process and architecture optimization.  Our approach extends to many other quantitative MRI applicationse.g. \cite{alexander2008,brihuegamoreno2003,grussu2020multi}, other imaging problems e.g. \cite{prevost2010}, and wider feature selection/experiment design problems e.g. \cite{marinescu2019,putten2000}.
 
\subsection*{Acknowledgements}

We thank: Tristan Clark and the HPC team (James O’Connor, Edward Martin); MUDI Organizers (Marco Pizzolato, Jana Hutter, Fan Zhang); Amy Chapman, Luca Franceschi, Marcela Konanova and Shinichi Tamura; NVIDIA for donating the GPU for winning 2019 MUDI. Funding details SB: EPRSC and Microsoft scholarship, EPSRC grants M020533 R006032 R014019, NIHR UCLH Biomedical Research Centre, HL: Research Initiation Project of Zhejiang Lab (No.2021ND0PI02), FG: Fellowships Programme Beatriu de Pinós (2020 BP 00117), Secretary of Universities and Research (Government of Catalonia).

%
%
\bibliographystyle{splncs03}
\bibliography{main}

\newpage

\section*{\centering \Large Supplementary Materials}

\begin{table}[ht]
\centering
\caption{
Ablation study of PROSUB's components (top-to-bottom) i) not computing an average measurement score, ii) not using RFE, iii) not progressively subsampling measurements during training (also fig.~\ref{removing_measurements_during_training:fig} right).  No NAS, across five random seeds / initializations on a single validation subject,  $ T_{1},T,M = 2,6,50 $.
}\label{ablation:table}
\begin{tabular}{c c c}
Formula Change & & Val MSE  \\ \hline
Alg.-line~\ref{average_measurement_score:line} is $ \overline{s}_{t} = s_{t}, \ s_{t} \in \mathbb{R}^{N} $ & & $ 3.02 \pm 0.01 $ \\ 
$ T_{1} = T = 6, \ D[T_{1}] = N-M $ & & $ 1.86 \pm 0.03 $ \\
Alg.-line~\ref{removing_measurements_during_training:line} is $ m^{e}_{t} = \max{\{ m_{t} - (e-E_{d})\mathbb{I}_{e \geq E_{d}}\cdot \mathbb{I}_{i \in D}(i), 0  \} } $  & & $ 1.52 \pm 0.03 $ \\
\hline
 PROSUB w/o NAS & & $ \textbf{1.47} \pm 0.08 $
\end{tabular}
\end{table}

\begin{table}[ht]
\centering
\caption{
All hyperparameters.  We conducted a brief search for the NAS hyperparameters, which ex. dropout are the same between PROSUB and SARDU-Net+NAS.  SARDU-Net w/o NAS baselines are from official implementation. 1st/2nd net. refers to respective fist/second network in the dual-network implementations. `choices in' means NAS can choose one of the following values e.g. \#Layers in 1st net = $ 2 \ choices \ in \  \{ 1,2,3 \} + 1 $ means the first layer in PROSUB $ S_{t} $ is replaced by $1,2,3$ layers. 
}\label{hyperparameters_all:table}
\resizebox{\linewidth}{!}{
\begin{tabular}{c c c c c c c c c c}
Hyerparameters & Value & & SARDU-Net & & SARDU-Net-NAS & & PROSUB \\
\hline
$ E $, Optimizer, Learning Rate & $ 200, ADAM, 1E-3 $ & & \checkmark & & \checkmark & & \checkmark \\
Batch Size, Weight Initialization & $ 1500 $, He Normal & & \checkmark & & \checkmark & & \checkmark \\
\multicolumn{1}{c}{\multirow{2}{*}{Data Normalization }} & Max-$99 \% $ & & \checkmark & & \checkmark & & \\
 & '' measurement-wise & &  & &  & & \checkmark  \\
\hline
$ T_{1}, T $ & $ 4, 36+9 $ & &   & & \checkmark & & \checkmark \\
$ \alpha_{t}, D_{t}, E_{d} $ & $ \frac{T-t}{T-1}, \frac{N-M}{T-T_{1}+1}, 20 $ & &   & &   & & \checkmark \\
\hline
NAS adapted from & AutoKeras w. Greedy (default) Strategy & &  & & \checkmark & & \checkmark \\
\multicolumn{1}{c}{\multirow{2}{*}{\#Layers in 1st/2nd net }} & $ 3 $ & & \checkmark & &   & & \\
 & $ 2 \ choices \ in \  \{ 1,2,3 \} + 1 $ & &  & & \checkmark  & & \checkmark  \\
 
\multicolumn{1}{c}{\multirow{2}{*}{\#Units in layer 1,2 1st net. }} & $ 1063, 781 \ (M=500) \ 417, 333 \ (M<500) $ & & \checkmark & &   & & \\
 & $ 2 $ choices in $ \{ 128, 256, ..., 2048 \} $ & &  & & \checkmark  & & \checkmark  \\
 
\multicolumn{1}{c}{\multirow{2}{*}{\#Units in layer 1,2 2nd net. }} & $ 781, 1063 $ & & \checkmark & &   & & \\
 & $ 2 $ choices in $ \{ 128, 256, ..., 2048 \} $ & &  & & \checkmark  & & \checkmark  \\
\multicolumn{1}{c}{\multirow{3}{*}{\ Dropout }} & $ 0.2 $ & & \checkmark & &   & & \\
  & $ 4 $ choices in $ \{ 0, 0.1, 0.2, 0.3, 0.4 \} $ & &  & &  \checkmark  & & \\
  & $ 0 $ & &   & &   & & \checkmark \\
\end{tabular}
}
\end{table}

\begin{table}[ht]
\centering
\caption{
PROSUB results with the same experimental settings as table~\ref{mudi_crossval_test:table}, at maximum network size of the NAS search space (see table~\ref{hyperparameters_all:table}).  Increasing the network capacity does not necessarily improve performance.
}\label{max_size:table}
\resizebox{\linewidth}{!}{
\begin{tabular}{c c c c c c c cc c c c c c c}
M=500 & & 250 & & 100 & & 50 & & 40 & & 30 & & 20 & & 10 \\
\hline
$ 1.38 \pm 0.14 $ & & $ 1.46 \pm 0.19 $ & & $ 1.79 \pm 0.13 $ & & $ 2.34 \pm 0.26 $ & & $ 2.61 \pm 0.38 $ & & $ 2.92 \pm 0.30 $ & & $ 3.39 \pm 0.33 $ & & $ 4.68 \pm 1.36 $ 

\end{tabular}
}
\end{table}

\begin{table}[ht]
\centering
\caption{
Architectures chosen by the AutoKeras Neural Architecture Search (NAS) in table~\ref{mudi_crossval_test:table} for respective network:  Selector/Scoring | Prediction/Reconstruction, omitting first and last $ N = 1344 $ units in both networks.  The (No-NAS) SARDU-Net and PROSUB w/o NAS have architectures $ N \rightarrow 1063 \rightarrow 781 \rightarrow N, \ N \rightarrow 781 \rightarrow 1063 \rightarrow N $, the SARDU-Net used a smaller selector network: $ N \rightarrow 417 \rightarrow 333 \rightarrow N $ for $ M < 500 $.
}\label{network_sizes:table}
\resizebox{\linewidth}{!}{
\begin{tabular}{c  c | c | c | c | c }
Split & Network & M=500 & 250 & 100 & 50  \\
\hline

\multirow{2}{*}{1} & SARDU-Net-NAS & 417,333 | 781,1063 & 417,333 | 781,1063 & 417\textsuperscript{2},333 | 781,1063 & 417,333 | 781,1063 \\ 
 & PROSUB & 1063,896 | 781\textsuperscript{2},1063 & 1063,896 | 781\textsuperscript{2},1063 & 1063,896 | 781\textsuperscript{2},1063 & 1063,512,896 | 781\textsuperscript{2},1063 \\

\multirow{2}{*}{2} & SARDU-Net-NAS & 417,333 | 781,1063 & 417,333\textsuperscript{2} | 781,1063 & 417,333 | 781,1063 & 417,333 | 781,1063 \\ 
 & PROSUB & 1063,896,781\textsuperscript{2} | 781,1063 & 1063,640 | 781,1063 & 1063,640 | 781,1063 & 640,128,640 | 781,1063 \\
 
\multirow{2}{*}{3} & SARDU-Net-NAS & 417,256 | 781,1063 & 417,256 | 781,1063 & 417\textsuperscript{2},256 | 781,1063 & 512,256 | 781,1063 \\  
 & PROSUB & 1063,2048 | 781,1063 & 1063,1536 | 781,1063 & 1063,2048 | 781,1063 & 1024,2048 | 781,1063 \\ 
 
\multirow{2}{*}{4} & SARDU-Net-NAS & 417,333 | 781,1063 & 417,333,333 | 781,1063 & 417,333 | 781,1063 & 417,1920,417,333 | 781,1063 \\
  & PROSUB & 1063,1024,781\textsuperscript{2} | 781\textsuperscript{2},1063 & 1063,1024,781\textsuperscript{2} | 781,1063  & 1063,1024,781,1920 | 781\textsuperscript{2},1063 & 1063\textsuperscript{3},1024,781\textsuperscript{2} | 781\textsuperscript{2},1063 \\ 
 
\multirow{2}{*}{5} & SARDU-Net-NAS & 417,1152 | 781,1063 & 417,333\textsuperscript{3} | 781,1063 & 417,333 | 781,1063 & 417\textsuperscript{2},333 | 781,1063 \\
 & PROSUB & 1063,781\textsuperscript{2} | 781,1063 & 1063,781\textsuperscript{2} | 781,1063 & 1063,781,896 | 781,1063 & 1063,1664,1152,781\textsuperscript{2} | 781,1063 \\ 
\\

 &  & M=40 & 30 & 20 & 10  \\ 
 \hline

\multirow{2}{*}{1} & SARDU-Net-NAS & 417,333 | 781,1063\textsuperscript{3} & 417,333 | 781,1063 & 417,333 | 781,1063 & 417,333 | 781,1063 \\ 
 & PROSUB & 1063,896 | 781,7781,1063 & 1063,1280,896 | 781\textsuperscript{2},1063 & 1063,384 | 781\textsuperscript{2},1063 & 1063,1408 | 781\textsuperscript{2},1063 \\
\multirow{2}{*}{2} & SARDU-Net-NAS & 417,333 | 781,1063\textsuperscript{2} & 1024,333 | 781,1063 & 417,333 | 781,1063 &  417,333 | 781,1063 \\
 & PROSUB & 1063,640 | 781,1063 & 1063,640 | 781\textsuperscript{3},1063 & 1063,640 | 781,1063 & 1063,640 | 781,1063 \\

\multirow{2}{*}{3} & SARDU-Net-NAS & 417,256 | 781,1063\textsuperscript{2} & 512,256 | 781,1063 & 417,333 | 781,1063 & 417,384 | 781,1063 \\ 
 & PROSUB & 1063,2048 | 781,1063 & 1063,2048 | 781,1063 & 1063,384 | 781,1063 & 1063,2048 | 781,1063 \\ 

\multirow{2}{*}{4} & SARDU-Net-NAS & 417,256 | 781,1063 & 1920,333 | 781,1063 & 417,333 | 781,1063 & 640,256,333 | 781,1063 \\ 
 & PROSUB & 1063,1024,781\textsuperscript{2} | 781,1063 & 1063,1024,781\textsuperscript{2}| 781\textsuperscript{2},1063 & 1063,1024,781\textsuperscript{2} | 781\textsuperscript{2},1063\textsuperscript{3} & 1063,1024,781\textsuperscript{2} | 781\textsuperscript{2},1063 \\

 \multirow{2}{*}{5} & SARDU-Net-NAS & 417,333 | 781,1063\textsuperscript{2} & 417,333 | 781,1063 & 417,333 | 781,1063 & 417,333 | 781,1063 \\
 & PROSUB  & 1063,781\textsuperscript{2} | 781,1063 & 1063,781\textsuperscript{2} | 781\textsuperscript{2},1063 & 1063,781\textsuperscript{2} | 781,1063 & 1063,781\textsuperscript{2} | 781,1063 

\end{tabular}
}
\end{table}

\end{document}